\documentclass[11pt, a4paper]{article}
\usepackage{styleBuding}
\usepackage{bm}
\usepackage{listings}
\usepackage[usenames,dvipsnames,svgnames,table]{xcolor}
\usepackage{amsmath}
\usepackage{amssymb}
\usepackage{graphicx}
\usepackage{slashed}
\usepackage{soul}
\usepackage{multirow}
\usepackage{subfigure}
\usepackage{siunitx} % for physics units
\usepackage[utf8]{inputenc}

\pdfoutput=1

\definecolor{back}{HTML}{F8F8F8}

\usepackage{epstopdf} % convert any .eps graphics to .pdf compiling with pdfLaTeX

\def\delew{\Delta_{EW}}

\def\to{\rightarrow}

\def\bi{\begin{itemize}}
\def\ei{\end{itemize}}

\def\sps1ap{SPS1a$^\prime$}
\def\c1p{C1$^\prime$}

\def\be{\begin{equation}}
\def\ee{\end{equation}}
\def\bea{\begin{eqnarray}}
\def\eea{\end{eqnarray}}
\def\beas{\begin{eqnarray*}}
\def\eeas{\end{eqnarray*}}

\newcommand\plb[3]{{\it Phys.\ Lett.\ }{\bf B #1} (#2) #3}

\begin{document}
\begin{titlepage}
%\begin{flushright}
%OU-HEP/121030
%\end{flushright}

\vspace{0.5cm}
\begin{center}
{\Large \bf Impact of the LZ Experiment on the DM Phenomenology and Naturalness in the MSSM
}\\
\vspace{1.2cm} \renewcommand{\thefootnote}{\fnsymbol{footnote}}
{\large Dongwei Li$^1$,
Lei Meng$^2$,
and Haijing Zhou$^2$\footnote[1]{Email: zhouhaijing0622@163.com }
}\\
\vspace{1.2cm} \renewcommand{\thefootnote}{\arabic{footnote}}
{\it
$^1$Fundamentals Department, Henan Police College, Zhengzhou 450046, China \\
}
{\it
$^2$School of Physics, Henan Normal University, Henan Xinxiang 453007, China \\
}

\end{center}

\vspace{0.5cm}
\begin{abstract}
\noindent
This paper uses approximate analytical formulas and numerical results with the bino-dominated dark matter (DM) as an example to  analyze the impact of the LUX-ZEPLIN (LZ) experiment on the DM phenomenology and naturalness in the Minimal Supersymmetric Standard Model (MSSM). We conclude that the limitation of the latest LZ experiment worsens the naturalness of the MSSM, as the predictions of the $Z$-boson mass and DM relic density demonstrate, particularly in the regions where the correct DM relic density is obtained by the $Z$- or $h$-mediated resonant annihilations.
\vspace*{0.8cm}
%\noindent PACS numbers: 12.60.Jv,14.80.Va,14.80.Ly

\end{abstract}

\end{titlepage}

\section{Introduction}

Supersymmetric models of particle physics are renowned for providing an elegant solution to the daunting
gauge hierarchy problem. The Minimal Supersymmetric Standard Model (MSSM), as the most economical supersymmetric expansion model, may provide a solid description of nature from the weak scale to energy scales associated with the grand unification~\cite{primer}, which also receives indirect experimental support from the measured strengths of weak-scale gauge couplings, measured value of the top quark mass, and discovery of an SM Higgs-like boson by ATLAS~\cite{atlas_h} and CMS~\cite{cms_h} in 2012. However, this audacious extrapolation has suffered a string of serious setbacks because
LHC and DM data have shown no signs of supersymmetric matter, which has led some physicists to question whether weak-scale SUSY really exists or at least to
concede that it suffers diversiform unattractive  fine tunings~\cite{shifman}.

In the MSSM, the Z-boson mass is as follows~\cite{wss}
\be
\frac{m_Z^2}{2} = \frac{m_{H_d}^2 + \Sigma_d^d - (m_{H_u}^2+\Sigma_u^u)\tan^2\beta}{\tan^2\beta -1} -\mu^2 \;,
\label{eq:mZs}
\ee
where $m_{H_u}^2$ and $m_{H_d}^2$ are the soft SUSY-breaking (not physical) Higgs mass terms;
$\mu$ is the superpotential higgsino mass term; $\tan\beta\equiv v_u/v_d$ is the ratio of Higgs field vevs;
$\Sigma_u^u$ and $\Sigma_d^d$ include various independent radiative corrections~\cite{rns}. Since the term ( $ m^2_{H_d}+{\Sigma_d^d}$ ) is suppressed by $\tan^2\beta -1$, for even moderate $\tan\beta$ values, Eq.~(\ref{eq:mZs}) approximately reduces to
\be
\frac{m_Z^2}{2} \simeq -(m_{H_u}^2+\Sigma_u^u)-\mu^2\;.
\label{eq:approx}
\ee
To naturally achieve $m_Z\simeq 91.2 {~\rm GeV}$, $-m_{H_u}^2$ , $-\mu^2$, and each contribution to
$-\Sigma_u^u$ should be  comparable in magnitude to $m_Z^2/2$. The extent of the comparability can be quantified using the {\rm electroweak fine tuning parameter}~\cite{Baer:2013bba}\footnote{Compared with the other two measures of EWFT ($\Delta_{BG}$ and $\Delta_{HS}$) as shown in Ref.~\cite{Baer:2013bba}, $\Delta_{EW}$ is created from weak-scale SUSY parameters and consequently contains no information about any possible high-scale origin. Hence, $\Delta_{EW}$ is advantageous because it is model-independent, where any model that yields the same weak-scale mass spectrum will generate the same value of $\Delta_{EW}$~\cite{Baer:2012cf,Baer:2013gva,Baer:2014ica}. Meanwhile, $\Delta_{EW}<\Delta_{BG}\lesssim\Delta_{HS}$, i.e., $\Delta_{EW}$ can be considered a lower bound on electroweak fine tuning~\cite{Baer:2013ava}. Any model with a large value of $\Delta_{EW}$ is always fine tuned.}
\be
\Delta_{EW} \equiv max_i \left|C_i\right|/(M_Z^2/2)\;,
\ee
where  $C_{H_u}=-m_{H_u}^2  $ , $C_\mu =-\mu^2$, and  $C_{\Sigma_u^u } =-\Sigma_u^u  $.
A lower value of $\delew$ implies less fine tuning, and $1/\Delta_{EW}$ is the percentage of fine tuning,
 {\it e.g.}, $\Delta_{EW}=20$ corresponds to $\Delta_{EW}^{-1}=5\%$ fine tuning among the terms that contribute to $m_Z^2/2$.
 Therefore, given the experimental lower bound, there is a general consensus that smaller values of $|\mu|$ are preferred in fine tuning issues. However, the current experiment limits impose a strong lower
  bound on $|\mu|$. For example, in 2017, the analysis of a global fit for the
MSSM, which considered various experimental constraints\footnote{These experimental constraints include those from the DM relic density, PandaX-II (2017) results for the SI cross section~\cite{PandaX-II:2017hlx}, PICO results for the SD cross section~\cite{PICO:2017tgi}, and searches for supersymmetric particles at the 13-TeV LHC with $36 {~\rm fb^{-1}}$ data (especially the CMS analysis of the
electroweakino production )~\cite{CMS:2018szt}. } showed that $\mu > 350{~\rm GeV}$ was favored at a $95\%$ confidence level (C.L.)~\cite{Bagnaschi:2017tru}. This value of $\mu$ can induce a tuning of approximately $3\%$ to predict the Z-boson mass.
  The studies in Ref.~\cite{Drees:2021pbh} showed that the Xenon-1T direct search limits~\cite{Aprile:2018dbl} imposed a strong lower bound on $|\mu|$, particularly for $\mu > 0$ or when the masses of the heavy Higgs bosons of the MSSM were near their current limit from LHC searches.
  Ref.~\cite{He:2023lgi} demonstrated that $\mu$ should be larger than approximately $500~{\rm GeV}$ for $M_1 < 0 $ and $630~{\rm GeV}$ for $M_1 > 100~{\rm GeV}$ considering the recent measurement of the muon anomalous magnetic moment at Fermilab~\cite{Abi:2021gix}, first results of the LUX-ZEPLIN (LZ) experiment in the direct search for DM ~\cite{LZ:2022ufs}, and rapid progress of the LHC search for supersymmetry\cite{CMS:2018szt,ATLAS:2021moa,CMS:2017moi}.
In Ref.~\cite{Arcadi:2022hve}, systematic studies on DM in the hMSSM with a light gaugino/higgsino sector
revealed that the stringent requirement of the conventional thermal paradigm as a mechanism to achieve the correct DM relic density
had a significant impact on the viable parameter space, one of which is that the lower bounds of $\mu$ were elevated to approximately
500 GeV.
   These improved bounds imply a tuning of ${\cal{O}}(1\%)$ to predict the $Z$-boson mass.

In this paper, using the approximate analytical formulas and numerical results,
we analyzed the DM phenomenology and associated unnaturalness in the MSSM in detail under the latest LZ experimental limits.
The rest of this paper is organized as follows. Section~\ref{sec:model} briefly introduces the neutralino sections of the MSSM and demonstrates the DM scattering cross-sections with nucleons and annihilation for bino-like $\tilde{\chi}^0_{1}$ using the approximate analytical formulas.
 Section~\ref{sec:scan} briefly describes our scanning strategy and investigates
  the predictions for the surviving samples and properties of bino-dominated DM scenarios to understand the associated unnaturalness.
  Section~\ref{sec:Conclusion} presents our conclusions.

\section{Dark Matter Section in the MSSM}
\label{sec:model}
In the MSSM, the neutralino mass matrix in the basis of $\Psi^0=(-i\tilde{B}^0,-i\tilde{W}^0,\tilde{H}_d^0,\tilde{H}_u^0)$ is~\cite{Pierce:2013rda}:

%\begin{equation}
%M_{neut}=\left(
%\begin{array}{cccc}
% M_1 & 0 & -\frac{v g_1c_{\beta}}{2} & \frac{v g_1 s_{\beta }}{2} \\
% 0 & M_2 & \frac{v c_W g_1 c_{\beta}}{2 s_W } & -\frac{v c_W g_1 s_{\beta }}{2 s_W } \\
% -\frac{v g_1c_{\beta}}{2} & \frac{v c_W g_1 c_{\beta}}{2 s_W } & 0 & -\mu  \\
% \frac{v g_1 s_{\beta }}{2 } & -\frac{v c_W g_1 s_{\beta }}{2 s_W } & -\mu  & 0 \\
%\end{array}
%\right)
%\end{equation}
%where $g_1= 2 M_Z s_W/v$, $s_\beta \equiv \sin\beta$ and $c_\beta\equiv \cos\beta$.
\begin{equation}
M_{\rm neut} = \left(
  \begin{array}{cccc}
    M_{1} & 0 & -c_{\beta}s_{W}m_{Z} & s_{\beta}s_{W}m_{Z}\\
        0 & M_{2} &c_{\beta}c_{W}m_{Z} & -s_{\beta}c_{W}m_{Z}\\
        -c_{\beta}s_{W}m_{Z} & c_{\beta}c_{W}m_{Z} & 0 & -\mu\\
        s_{\beta}s_{W}m_{Z} & - s_{\beta}c_{W}m_{Z} & -\mu & 0
  \end{array}
  \right ),
\end{equation}
where $M_{1},\ M_{2}$, and $\mu$ are the soft SUSY-breaking mass parameters of the bino, wino, and higgsinos, respectively;
 $m_{\rm Z}$ is the $Z$-boson mass; $\theta_{\rm w}$ is the Weinberg angle ($c_{W} \equiv \cos\theta_{W}$ and $s_{W} \equiv \sin\theta_{W}$);
$\tan\beta\equiv s_\beta/c_\beta=v_{u}/v_{d}$ is the ratio of the vacuum expectation values for the two Higgs doublets ($c_\beta \equiv \cos\beta$ and $s_\beta \equiv \sin\beta$) and $v^2=v_{u}^2+v_{d}^2=(246{~\rm GeV})^2$.
Diagonalizing $M_{\rm neut}$ with a $4\times4$ unitary matrix N yields the masses of the physical states $\tilde\chi^0_i$ (ordered by mass) of four neutralinos:
\begin{eqnarray}
 N^* M_{\rm neut} N^{-1}  = {\rm diag} \{ m_{\tilde\chi^0_1}, m_{\tilde\chi_2^0}, m_{\tilde\chi_3^0}, m_{\tilde\chi_4^0} \} \nonumber
\end{eqnarray}
with
\begin{eqnarray}\label{2}
    \tilde\chi^0_i = N_{i1}\tilde B^0 + N_{i2}\tilde W^0 + N_{i3}\tilde H^0_d + N_{i4}\tilde H^0_u  ~~~~(i=1,2,3,4), \nonumber
\end{eqnarray}
where $m_{\tilde\chi^0_i}$ is the root to the following eigenequation:
\begin{eqnarray}
\left(x-M_1\right) \left(x-M_2\right)  (x^2-\mu^2 )-m_Z^2 \left(x-M_1c_W^2-M_2 s_W^2\right) \left(2 \mu  s_{\beta }c_{\beta}+x\right) =0 .
\end{eqnarray}
The eigenvector of $m_{\tilde\chi^0_i}$ is the column vector constituted by $N_{ij}(j=1,2,3,4)$, which is given by
\begin{equation}\label{4}
N_i =\frac{1}{\sqrt{C_i}}\left(\begin{array}{c}
(\mu^2 - m^2_{\tilde\chi^0_i})(M_2- m_{\tilde\chi^0_i})s_W \\
-(\mu^2 - m^2_{\tilde\chi^0_i})(M_1- m_{\tilde\chi^0_i})c_W  \\
 (M_2 s^2_W + M_1 c^2_W -m_{\tilde\chi^0_i}) (m_{\tilde\chi^0_i}c_\beta + \mu s_\beta)m_Z \\
-(M_2 s^2_W + M_1 c^2_W -m_{\tilde\chi^0_i}) (m_{\tilde\chi^0_i}s_\beta + \mu c_\beta)m_Z\\
\end{array}\right).
\end{equation}
The specific form of the normalization factor $C_i$ is:
\begin{eqnarray}\label{5}
  C_i &=& (\mu^2 - m^2_{\tilde\chi^0_i})^2 (M_2- m_{\tilde\chi^0_i})^2 s_W^2
  + (\mu^2 - m^2_{\tilde\chi^0_i})^2 (M_1- m_{\tilde\chi^0_i})^2 c_W^2  \nonumber\\
  && + (M_2 s_W^2 + M_1 c_W^2 - m_{\tilde\chi^0_i})^2 (\mu^2 + m^2_{\tilde\chi^0_i} + 4\mu m_{\tilde\chi^0_i}s_\beta c_\beta)m_Z^2.
\end{eqnarray}
Then, the diagonalizing matrix is $N=\{N_1,N_2,N_3,N_4\}$, where ${i=1,2,3,4}$ denotes the $i$-th neutralino.

This work focuses on the lightest neutralino, $\tilde\chi^0_1$, which acts as the DM candidate. $N_{11}^2$, $N_{12}^2$, and $N_{13}^2+N_{14}^2$
 are the bino, wino, and higgsino components
in the physical state $\tilde\chi^0_1$, respectively, and satisfy  $N_{11}^2+N_{12}^2 +N_{13}^2+N_{14}^2=1$. If $N_{11}^2 > 0.5 $($N_{12}^2>0.5$ or $N_{13}^2+N_{14}^2>0.5$), we call $\tilde\chi^0_1$  the bino- ( wino- or higgsino- ) dominant DM.
The couplings of DM to the scalar Higgs states and Z-boson are included in the calculation of the DM-nucleon cross sections and DM annihilation, which correspond to the Lagrangian\cite{MSSM1,MSSM2}:
\begin{eqnarray}\label{6}
  {\cal L}_{MSSM}\ni C_{\tilde\chi^0_1 \tilde\chi^0_1 h} h \bar{\tilde\chi}^0_1 \tilde\chi^0_1
  +  C_{\tilde\chi^0_1 \tilde\chi^0_1 H} H \bar{\tilde\chi}^0_1 \tilde\chi^0_1
  +C_{\tilde\chi^0_1 \tilde\chi^0_1 Z}Z_\mu \bar{\tilde\chi}^0_1 \gamma^\mu\gamma_5 \tilde\chi^0_1.  \nonumber
\end{eqnarray}
The coefficients are:
\begin{equation}\label{xxh}
    C_{\tilde\chi^0_1 \tilde\chi^0_1 h} \approx \frac{2m_Z^2\mu}{v C_1}(\mu^2 -m^2_{\tilde\chi^0_1})(M_2 s^2_W + M_1 c^2_W - m_{\tilde\chi^0_1})^2 (\frac{m_{\tilde\chi^0_1}}{\mu}+\sin2\beta),
\end{equation}
\begin{equation}\label{xxH}
    C_{\tilde\chi^0_1 \tilde\chi^0_1 H} \approx \frac{2m_Z^2\mu}{v C_1}(\mu^2 -m^2_{\tilde\chi^0_1})(M_2 s^2_W + M_1 c^2_W - m_{\tilde\chi^0_1})^2 \cos2\beta,
\end{equation}
\begin{equation}\label{xxz}
    C_{\tilde\chi^0_1 \tilde\chi^0_1 Z} \approx \frac{ m_Z^3}{v C_1}(\mu^2 -m^2_{\tilde\chi^0_1})(M_2 s^2_W + M_1 c^2_W - m_{\tilde\chi^0_1})^2 \cos2\beta,
\end{equation}
where $h$ and $H$ are two CP-even Higgs states predicted by the MSSM: the SM-like Higgs boson and non-SM doublet Higgs boson, respectively.
%$\alpha$ is their  mixing angle of the CP-even Higgs states satisfying $\alpha\simeq\beta-\pi/2$|A in the large CP-odd Higgs mass ($m_A$) limit [95].

Serving as a weakly interacting massive particle (WIMP), $\tilde\chi^0_1$ may be detected by measuring their spin-independent (SI) and spin-dependent (SD) scattering cross-sections after an elastic scattering of $\tilde\chi^0_1$ on a nucleus occurs. At the tree level, the contribution to the SD (SI) scattering cross-section in the heavy squark limit is dominated by the t-channel Z-boson (CP-even Higgs bosons $h_i$) exchange diagram. Therefore, the scattering cross-sections have the following form \cite{He:2023lgi,Baum:2021qzx,Baum:2023inl}:
\begin{equation}\label{sd0}
   \sigma^{SD}_{\tilde\chi^0_1 -N}\approx C_N\times(\frac{ C_{\tilde\chi^0_1 \tilde\chi^0_1 Z}}{0.01})^2,
\end{equation}
\begin{eqnarray}\label{si0}
   \sigma^{SI}_{\tilde\chi^0_1 -N}& = & \frac{m_N^2}{\pi v^2}
   \Big(\frac{m_N m_{\tilde\chi^0_1}}{m_N+m_{\tilde\chi^0_1}}\Big)^2
   \Big(\frac{1}{125 GeV}\Big)^4 \times \nonumber \\
   &&\Big\{ {(F_u^N +F_d^N ) C_{\tilde\chi^0_1 \tilde\chi^0_1 h}}\Big(\frac{125 GeV}{m_h}\Big)^2 + {(\frac{F_u^N}{\tan\beta}- F_d^N \tan\beta) C_{\tilde\chi^0_1 \tilde\chi^0_1 H}}\Big(\frac{125 GeV}{m_H}\Big)^2\Big\}^2 \nonumber \\
   &\approx & 6.4\times 10^{-44} cm^2 \times \left ( \frac{F_u^N + F_d^N}{0.28} \right )^2 \times  \nonumber \\
   && \left\{ \left ( \frac{F_u^N}{F_u^N + F_d^N} \right ) \times  \left[\frac{C_{\tilde\chi^0_1 \tilde\chi^0_1 h}}{0.1}\Big(\frac{125 GeV}{m_h}\Big)^2
   +\frac{1}{\tan\beta}\frac{C_{\tilde\chi^0_1 \tilde\chi^0_1 H}}{0.1}\Big(\frac{125 GeV}{m_H}\Big)^2\right] \right . \nonumber \\
    &&\left .  + \left ( \frac{F_d^N}{F_u^N + F_d^N} \right )\times \left[{\frac{C_{\tilde\chi^0_1 \tilde\chi^0_1 h}}{0.1}\Big(\frac{125 GeV}{m_h}\Big)^2-\tan\beta \frac{C_{\tilde\chi^0_1 \tilde\chi^0_1 H}}{0.1}\Big(\frac{125 GeV}{m_H}\Big)^2}\right] \right\}^2,
\end{eqnarray}
where N=p, n represents the proton and neutron, and $C_p \simeq 2.9 \times 10^{-41}~{\rm cm^2} $ ($C_n \simeq 2.3 \times 10^{-41}~{\rm cm^2} $)~\cite{Badziak:2015exr,Badziak:2017uto}.
The form factors at zero momentum transfer are $F^{(N)}_d=f^{(N)}_d+f^{(N)}_s+\frac{2}{27}f^{(N)}_G$ and $F^{(N)}_u=f^{(N)}_u+\frac{4}{27}f^{(N)}_G$, where $f^{(N)}_q =m_N^{-1}\left<N|m_qq\bar{q}|N\right> $ ($q=u,d,s$)
is the normalized light quark contribution to the nucleon mass,
and $f^{(N)}_G=1-\sum_{q=u,d,s}f^{(N)}_q$ affects other heavy quark mass fractions in the nucleons~\cite{Drees1993,Drees1992}.
In this study, the default settings for $f_q^{N}$ were used in the micrOMEGAs package~\cite{Belanger2008}, and they predicted $F_u^{p} \simeq F_u^n \simeq 0.15$ and $F_d^{p} \simeq F_d^n \simeq 0.13$.
Hence, the DM-proton scattering and DM-neutron scattering had approximately equal SI cross-sections (i.e., $\sigma^{SI}_{{\tilde {\chi}^0_1}-p} \simeq \sigma^{SI}_{{\tilde {\chi}^0_1}-n}$)~\cite{DM-detecion-SI-SD}.

In the pure bino limit ($m_{\tilde\chi^0_1} \approx M_1 $ and $N_{11}^2 \approx 1$),  the above formulas can be approximated as
\begin{equation}\label{10}
    C_1=(\mu^2-m^2_{\tilde\chi^0_1})^2(m_{\tilde\chi^0_1}-M_2)^2 s_W^2,
\end{equation}
\begin{equation}\label{bino_xxh}
    C_{\tilde\chi^0_1 \tilde\chi^0_1 h} \approx \frac{2m_Z^2\mu s^2_W}{v (\mu^2 -m^2_{\tilde\chi^0_1})} (\frac{m_{\tilde\chi^0_1}}{\mu}+\sin2\beta),
\end{equation}
\begin{equation}\label{bino_xxH}
    C_{\tilde\chi^0_1 \tilde\chi^0_1 H} \approx \frac{2m_Z^2\mu s^2_W}{v (\mu^2 -m^2_{\tilde\chi^0_1})} \cos2\beta,
\end{equation}
\begin{equation}\label{bino_xxz}
    C_{\tilde\chi^0_1 \tilde\chi^0_1 Z} \approx \frac{ m_Z^3 s^2_W }{v (\mu^2 -m^2_{\tilde\chi^0_1})} \cos2\beta.
\end{equation}
Meanwhile, taking $F^{(N)}_d \simeq F^{(N)}_u \simeq 0.14 $ ~\cite{Huang:2014xua} and $\tan \beta \gg 1$, we can conclude that
\begin{equation}\label{sd1}
   \sigma^{SD}_{\tilde\chi^0_1 -n}\approx 2.4 \times 10^{-40} cm^2\times\frac{m^2_Z v^2}{\mu^4}\times(\frac{1}{1-m_{\tilde\chi^0_1}^2/\mu^2})^2,
\end{equation}
\begin{eqnarray}\label{si1}
   \sigma^{SI}_{\tilde\chi^0_1 -N}
   &\approx & 2.1\times 10^{-45} cm^2 \times \frac{v^2}{\mu^2} \times (\frac{ 1 }{1-m_{\tilde\chi^0_1}^2/\mu^2})^2 \nonumber \\
   && \times \Big\{ (\frac{m_{\tilde\chi^0_1}}{\mu}+{\sin2\beta})\Big(\frac{125 GeV}{m_h}\Big)^2  + \frac{\tan\beta}{2} \Big(\frac{125 GeV}{m_H}\Big)^2\Big\}^2.
\end{eqnarray}
These two analytic formulas suggest that $\sigma^{SD} $ and $\sigma^{SI}$ are suppressed by $\mu^4$ and $\mu^2$, respectively.
Moreover, if $(\frac{m_{\tilde\chi^0_1}}{\mu}+{\sin2\beta})$ and $\tan\beta $ have opposite signs,
the contributions to $\sigma^{SI}$ from the light Higgs($h$) and heavy Higgs($H$) exchange channels destructively interfere with each other.
$\sigma^{SI}$  will vanish for $(\frac{m_{\tilde\chi^0_1}}{\mu}+{\sin2\beta})\Big(\frac{1}{m_h}\Big)^2  + \frac{\tan\beta}{2} \Big(\frac{1}{m_H}\Big)^2 \rightarrow 0$,
which is known as the ``generalized blind spot''~\cite{Baum:2023inl,Huang:2014xua}. For the convenience of subsequent descriptions,
we defined ${\cal{A}}_h= (\frac{m_{\tilde\chi^0_1}}{\mu}+{\sin2\beta})\Big(\frac{125 \rm GeV}{m_h}\Big)^2$
and ${\cal{A}}_H= \frac{\tan\beta}{2} \Big(\frac{125 {\rm GeV}}{m_H}\Big)^2$ to represent the contributions from  h and H to the
SI cross-section, respectively.

For DM that was produced via the standard thermal freeze-out, the relic density at freeze-out temperature $T_F \equiv m_{\tilde\chi^0_1}/x_F$ (typically, $x_F \simeq 20$) was approximately~\cite{Baum:2023inl}
\begin{equation} \label{eq:relic_density}
    \Omega_{\tilde\chi^0_1} h^2 \sim 0.12 \times \frac{2.5 \times 10^{-26} {\rm cm}^3/{\rm s}}{\left\langle \sigma v \right\rangle_{x_F}} \;,
\end{equation}
where $\left\langle \sigma v \right\rangle_{x_F}$ corresponds to the effective (thermally averaged) annihilation cross-section.
To match the observed DM relic density ($\Omega_{\rm DM} h^2 \simeq 0.12$)~\cite{Planck:2018vyg},
$\left\langle \sigma v \right\rangle_{x_F}\sim 2.5 \times 10^{-26} {\rm cm}^3/{\rm s}$ is required.

In the MSSM, for $ m_{\tilde\chi^0_1} < 1 {~\rm TeV}$, only the bino-dominated ${\tilde\chi^0_1}$  can predict the correct DM relic density. For the higgsino- or wino-dominated ${\tilde\chi^0_1}$,   the predicted relic density is much smaller than the observed DM relic density because they relatively strongly interacted with the Standard Model particles~\cite{Cao:2019qng}.
 In our scenario, considering bino-like ${\tilde\chi^0_1}$ as an example, we will discuss the fine tunings introduced by the DM sector in the MSSM under the current DM experimental limits.
For bino-like ${\tilde\chi^0_1} $, the contributions to $\left\langle \sigma v \right\rangle_{x_F}$ are from two channels
\begin{itemize}
  \item  The Z- or h-mediated resonant annihilation ~\cite{Han:2013gba,Cabrera:2016wwr}. The corresponding annihilation cross-sections are approximated to~\cite{Baum:2017enm}
   \begin{equation}\label{Z-funnel0}
    \langle\sigma v\rangle^{d\bar d,Z}_{x_F}\sim
    (2.5\times10^{-26}\frac{{\rm cm}^3}{s})\times\left[(\frac{0.46m_d}{1{\rm GeV}})\times\frac{ C_{\tilde\chi^0_1 \tilde\chi^0_1 Z}}{(1-\frac{m^2_Z}{4m^2_{\tilde\chi^0_1}})}\right]^2\left(\frac{m_{\tilde\chi^0_1}}{46{\rm GeV}}\right)^{-4},
   \end{equation}
   \begin{equation}\label{h-funnel0}
    \langle\sigma v\rangle^{d\bar d,h}_{x_F}\sim
    (2.5\times10^{-26}\frac{{\rm cm}^3}{s})\times\left[(\frac{0.048m_d}{1 {\rm GeV}})\times\frac{ C_{\tilde\chi^0_1 \tilde\chi^0_1 h}}{(1-\frac{m^2_h}{4m^2_{\tilde\chi^0_1}})}\right]^2\left(\frac{m_{\tilde\chi^0_1}}{62{\rm GeV}}\right)^{-2}.
\end{equation}
Due to the hierarchy of Yukawa couplings,
 the contribution to the thermal cross-section from the (${\tilde{\chi}_1^0} {\tilde{\chi}_1^0}\to q\bar{q}$) annihilations will be dominated by bottom quarks for the lighter $m_{\tilde\chi^0_1}$.
 Here, ${m_d}$ is the down-type quark mass. As the expression shows, the measured relic density requires a high degeneracy between $m_{Z(h)}$ and $2 m_{\tilde\chi^0_1}$ .
 We define degeneracy parameters $\Delta_{Z(h)}=|1-\frac{m^2_{Z(h)}}{4m^2_{\tilde\chi^0_1}}|$ to quantize the fine tuning between $m_{Z(h)}$ and $2m_{\tilde\chi^0_1}$, e.g.,
 $\Delta_{Z} = 10^{-3}$ implies $ 0.1\%$ fine tuning among $m_Z$ and $2m_ {\tilde\chi^0_1}$.

  \item  Co-annihilation  with sleptons, wino-dominated neutralinos, and/or charginos
  ~\cite{He:2023lgi,Baum:2023inl,Han:2013gba,Cabrera:2016wwr,Ellis:1998kh,Ellis:1999mm,Buckley:2013sca,Baker:2018uox, Yanagida:2019evh}.  The corresponding reactions are $\tilde{\chi}_i \tilde{\chi}_j \rightarrow X X^\prime $, where $X$ and $X^\prime$ donate SM particles. In this case, $\langle \sigma v\rangle_{x_F}$ in Eq.~(\ref{eq:relic_density}) should be replaced by $\langle \sigma_{eff} v\rangle_{x_F}$, and $\sigma_{eff} $ is given by~\cite{Griest:1990kh}
\begin{eqnarray}
    \sigma_{eff} &=& \sum_{i,j}\sigma_{ij}\frac{g_i g_j}{g^2_{eff}}(1+\Delta_i)^{3/2}(1+\Delta_j)^{3/2}\times exp[-x(\Delta_i+\Delta_j)], \label{Co-annihilation experession}
\end{eqnarray}
where
\begin{eqnarray}
\sigma_{ij}=\sigma(\tilde{\chi}_i \tilde{\chi}_j \rightarrow X X^\prime ), \nonumber
\end{eqnarray}
\begin{eqnarray}
g_{eff} \equiv   \sum_{i} g_i(1+\Delta_i)^{3/2}exp(-x\Delta_i), \nonumber
\end{eqnarray}

\begin{eqnarray}
\Delta_i \equiv (m_i - m_{\tilde{\chi}_1^0})/m_{\tilde{\chi}_1^0} ,\label{delta} \nonumber
\end{eqnarray}
$x \equiv m_{\tilde{\chi}_1^0}/T$. $g_i$ represents the internal degrees of freedom, and $\Delta_i$  parameterizes the mass splitting. In the co-annihilation case,  the right relic abundance usually  requires holding $\Delta_i$ at the $5-15\%$ level~\cite{Griest:1990kh}, which also corresponds to one  fine tuning.

\end{itemize}

\section{Numerical Results and Theoretical Analysis}
\label{sec:scan}
We used the \textsf{MultiNest} algorithm~\cite{Feroz:2008xx} with $n_{\rm live} = 10000$\footnote{The $n_{\rm live}$ parameter in the algorithm controls the number of active points sampled in each iteration of the scan.} to comprehensively scan the following parameter space:
\begin{eqnarray}\label{scan}
&& 1\leq\tan\beta\leq 60,\quad 0.1{\rm ~TeV} \leq \mu \leq 1 {\rm ~TeV},\quad 0.5{\rm ~TeV}\leq M_{A} \leq 10 {\rm ~TeV} ,
\nonumber\\
&& -1.0{\rm ~TeV}\leq M_1 \leq 1.0 {\rm ~TeV}, \quad 0.1{\rm ~TeV}\leq M_2 \leq 1.5 {\rm ~TeV}, \nonumber\\
&& -5 {\rm ~TeV} \leq A_{t}= A_{b}\leq 5 {\rm ~TeV}, \quad  0.1{\rm ~TeV}\leq M_{\tilde{\mu}_L}, M_{\tilde{\mu}_R}\leq 1 {\rm ~TeV}, \nonumber
\end{eqnarray}
 where $\tan \beta$ was defined at the electroweak scale, and the others were defined at the renormalization scale $Q=1~{\rm TeV}$. To obtain the SM-like Higgs boson mass ($m_h\approx125 ~\rm GeV$), the soft trilinear coefficients $A_t$ and $A_b$ were assumed to be equal and freely change to adjust the Higgs mass spectrum.
The masses of the second-generation sleptons ($ M_{\tilde{\mu}_L}, M_{\tilde{\mu}_R}$) were used as free parameters to explain the muon g-2 anomaly and predict the measured DM relic abundance by co-annihilation with sleptons. Other SUSY parameters of sleptons were fixed at 3 TeV to reduce the number of free parameters. The release of $\tilde{\tau}$ may change the $e/\mu$ signals of this study and relax the LHC restrictions ~\cite{Hagiwara:2017lse,Chakraborti:2023pis}. Other unimportant parameters were also fixed at 3 TeV, including the gluino mass $M_3$ and three generations of squarks except $A_t$ and $A_b$.

During the scan, some experimental constraints were imposed by constructing the following corresponding likelihood function to guide the
process as follows:
\begin{eqnarray}
\mathcal{L}=\mathcal{L}_{{{h}}} \times \mathcal{L}_{h, {\rm extra}} \times {\mathcal{L}_{B} }   \times  {\mathcal{L}_{\Omega h^2} }\times  {\mathcal{L}_{DD}} \times  {\mathcal{L}_{Vac}}  \times \mathcal{L}_{a_\mu} ,
\label{Likelihood}
\end{eqnarray}
where
\begin{itemize}
  \item $\mathcal{L}_{{{h}}}$ and $\mathcal{L}_{h, {\rm extra }}$ are the likelihood functions for the consistency of $h$'s properties with the LHC Higgs data at the $95\%$ C.L.~\cite{HS2020uwn} and collider searches for extra Higgs bosons~\cite{HB2020pkv}. The two restrictions were implemented by the programs \textsf{HiggsSignal\,2.6.2}~\cite{HS2013xfa,HSConstraining2013hwa,HS2014ewa,HS2020uwn} and ~\textsf{HiggsBounds\,5.10.2}~\cite{HB2008jh,HB2011sb,HB2013wla,HB2020pkv} , respectively.
  \item  ${\mathcal{L}_{B} }$ is the likelihood function for the measured  branching ratio of the $B \to X_s \gamma$ and $B_s \to \mu^+\mu^-$. These ratios were calculated by the formulae in Refs.~\cite{Domingo:2007dx,Domingo:2015wyn} and should be consistent with their experimental measurements at the $2\sigma$ level~\cite{Tanabashi:2018oca}.
  \item ${\mathcal{L}_{\Omega h^2} }$ is the likelihood function for the DM relic density with the central value of $0.120$ from the Planck-2018 data~\cite{Planck:2018vyg}. We assumed  theoretical uncertainties of $20\%$ in the density calculation. ${\mathcal{L}_{DD}}$ is the likelihood function for the $90\%$ C.L. upper bounds of the PandaX-4T experiment on the SI DM-nucleon scattering~\cite{PandaX-4T:2021bab} and XENON-1T experiment on the SD scattering~\cite{Aprile:2019dbj}. These DM observables were calculated using the package \textsf{MicrOMEGAs\,5.0.4}~\cite{Belanger:2001fz, Belanger:2005kh,Belanger:2004yn,Belanger:2006is, Belanger:2010pz, Belanger:2013oya, Barducci:2016pcb,Belanger:2018ccd}.
  \item ${\mathcal{L}_{Vac}}$ is the likelihood function for the vacuum stability of the scalar potential, which consists of the Higgs fields and the last two generations of the slepton fields. This condition was implemented by the code \textsf{Vevacious}~\cite{Camargo-Molina:2013qva,Camargo-Molina:2014pwa}.
  \item $\mathcal{L}_{a_\mu}$ is the likelihood function of the muon $g-2$ anomaly given by
\begin{eqnarray}
\mathcal{L}_{a_\mu} \equiv Exp\left[-\frac{1}{2} \left( \frac{a_{\mu}^{\rm SUSY}- \Delta a_\mu}{\delta a_\mu }\right)^2\right] = Exp\left[-\frac{1}{2} \left( \frac{a_{\mu}^{\rm SUSY}- 2.51\times 10^{-9}}{5.9\times 10^{-10} }\right)^2\right], \nonumber
\end{eqnarray}
where $\Delta a_\mu \equiv a_\mu^{\rm Exp} - a_\mu^{\rm SM}$ is the difference between experimental central value of $a_\mu$ and its SM prediction, and $\delta a_\mu$ is the total uncertainties in determining $\Delta a_\mu$~\cite{Bennett:2006fi,Abi:2021gix,Aoyama:2020ynm}.
\end{itemize}
We defined $\mathcal{L} = 1$ if the restrictions were satisfied; otherwise, $\mathcal{L} = Exp[-100]$.

To probe into the impact of the LZ experiment on the DM phenomenology and naturalness of the complete parameter space in the MSSM, we did not consider the constraints from the LHC search for SUSY. Similar to the studies in Ref.~\cite{He:2023lgi}, the restrictions from the LHC experiment require that the lower bounds of $\mu$ become approximately $400~{\rm GeV}$, and the upper bounds of $M_1$ become approximately $570~{\rm GeV}$, which severely compresses the surviving space of the MSSM. For example, compared with the restriction from the PandaX-4T experiment on the SI scattering cross-section, the lower bounds of $\mu$ improve by approximately $100~{\rm GeV}$. As a result, the regions with the $Z$- and $h$-mediated resonant annihilations are more unnatural at predicting the correct relic density so that they are commonly missed in the scans by the \textsf{MultiNest} algorithm.

\begin{figure}[t]
	\centering
    %\vspace{-0.2cm}
	\includegraphics[width=0.5\textwidth]{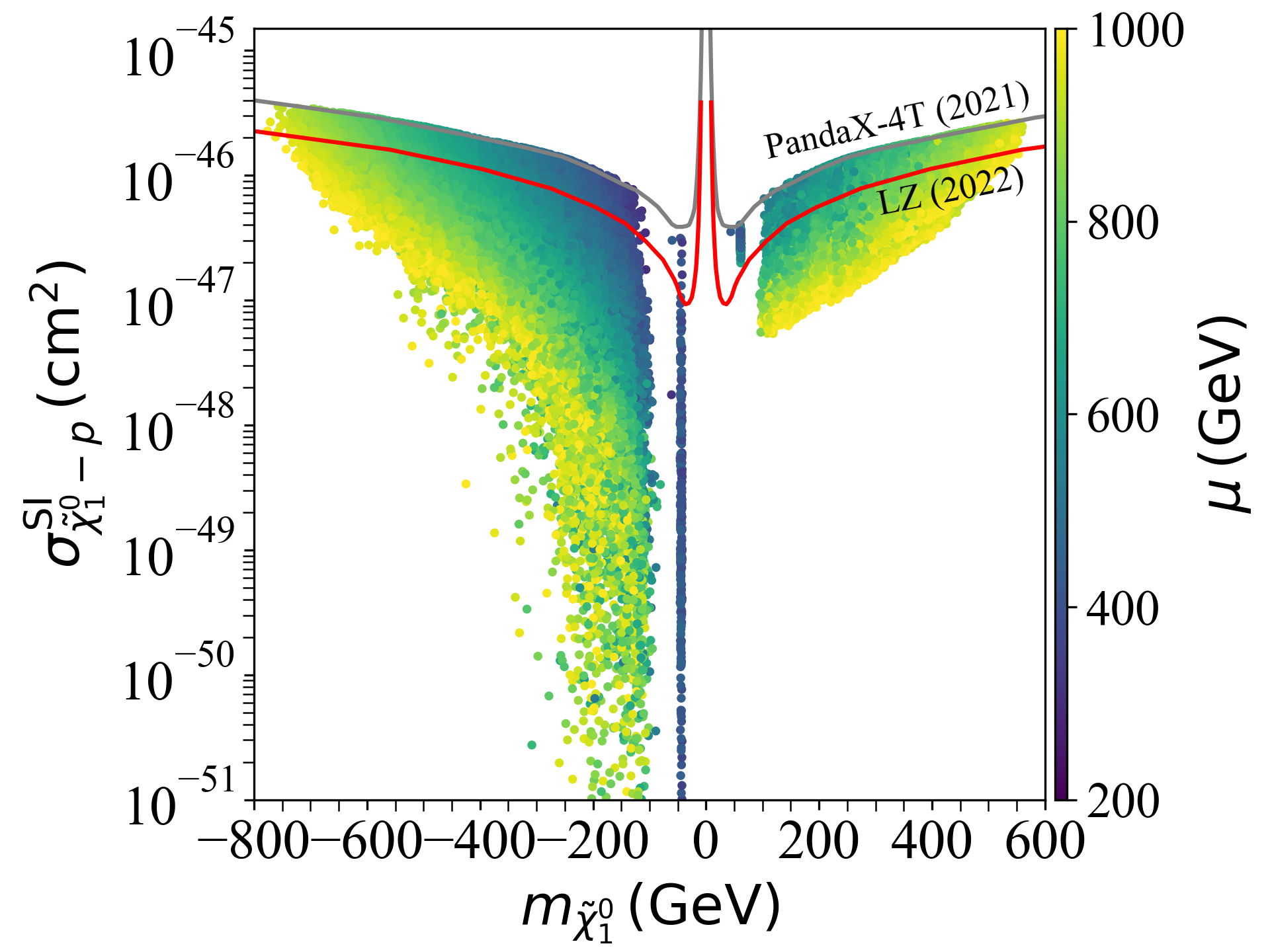}%\hspace{-0.3cm}
	\includegraphics[width=0.5\textwidth]{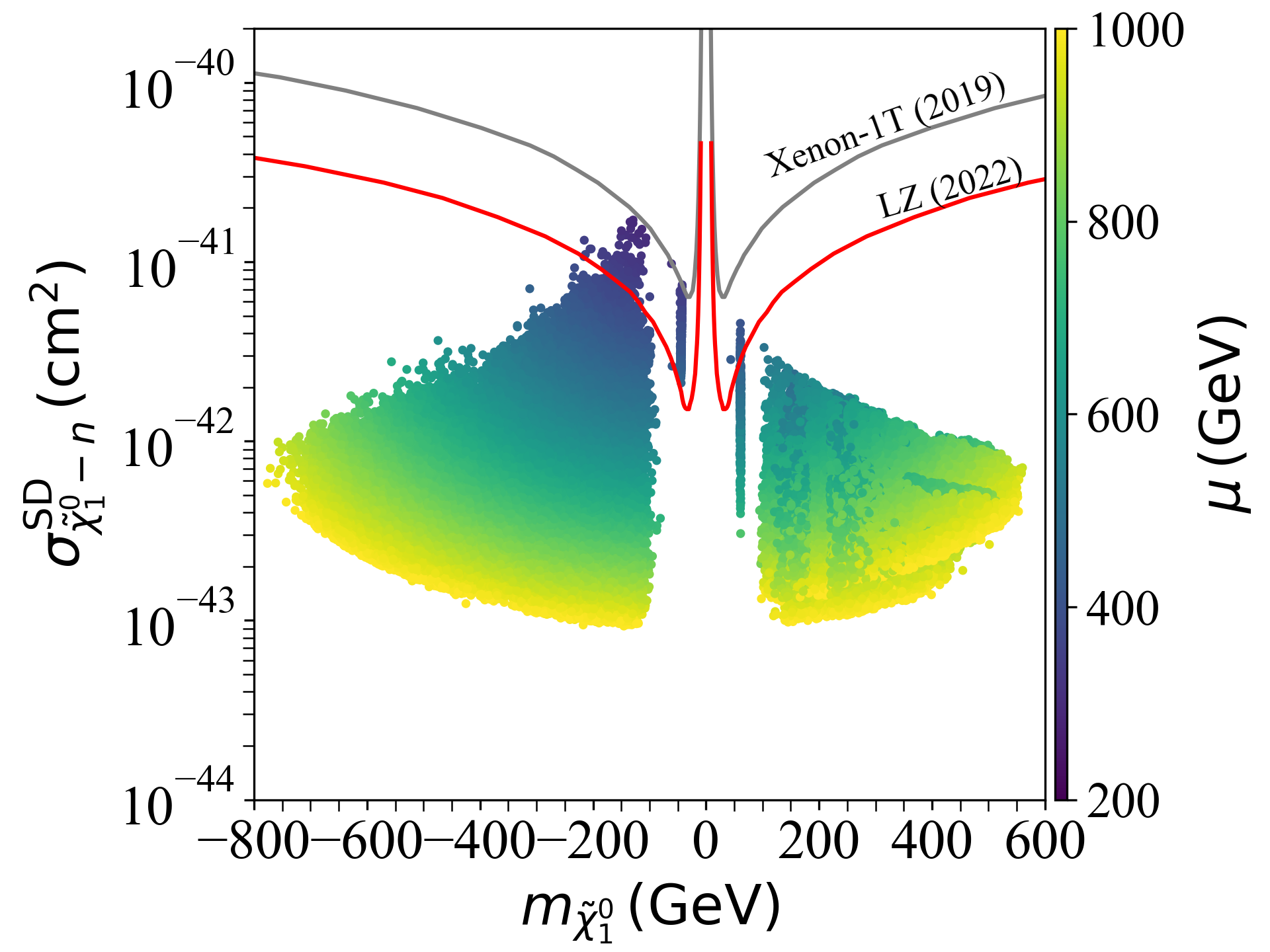}%\hspace{-0.4cm}

	\caption{\label{fig1}
		Projection of the refined samples onto
$ m_{\tilde\chi^0_1}-\sigma^{SI}_{\tilde\chi^0_1 - p}$(left panel) and
$m_{\tilde\chi^0_1}-\sigma^{SD}_{\tilde\chi^0_1 -n}$(right panel). The colors indicate the value of the higgsino mass $\mu$.  }
\end{figure}

\begin{figure}[h]
	\centering
%\vspace{-0.2cm}
	\includegraphics[width=0.5\textwidth]{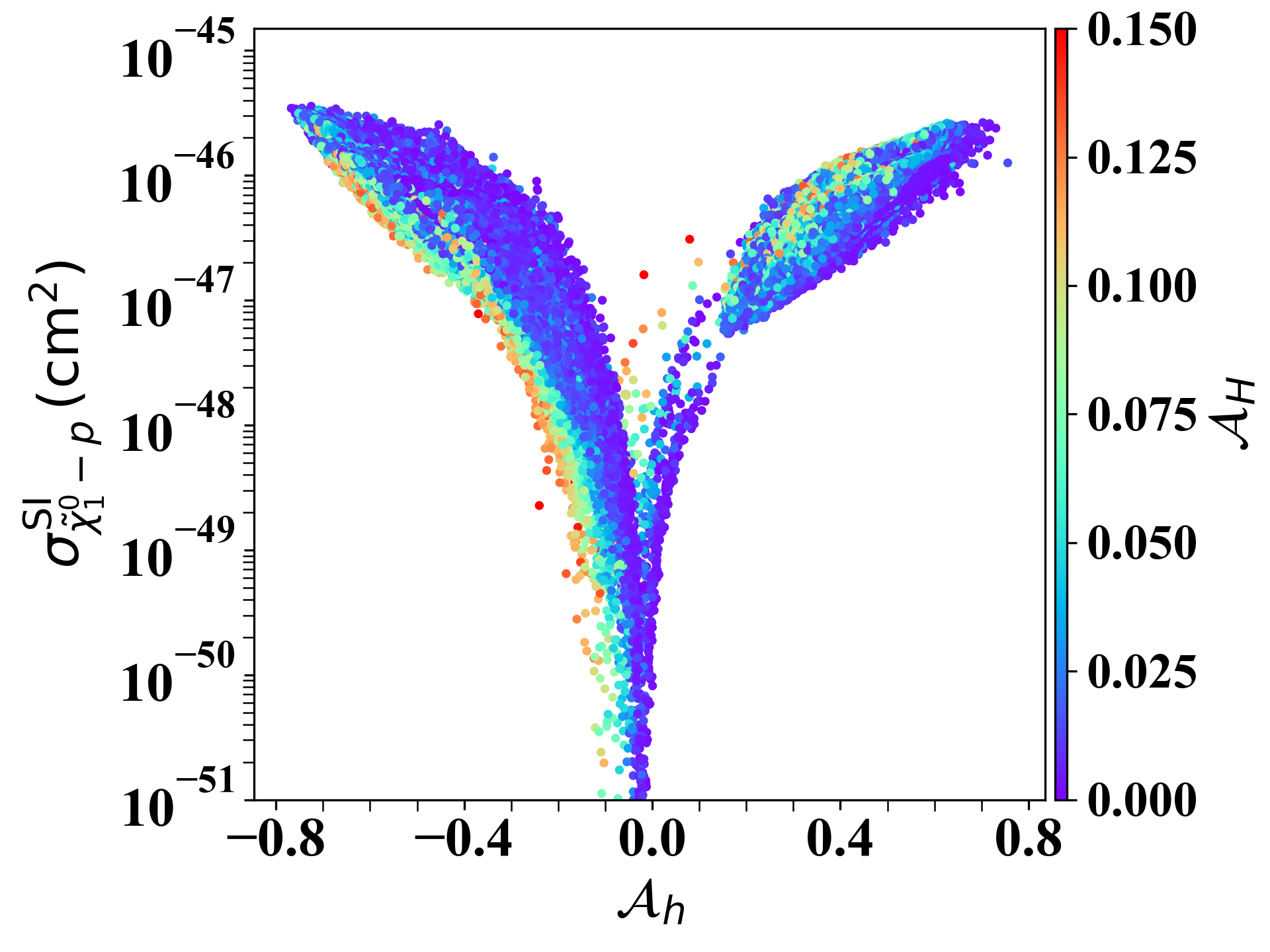}%\hspace{-0.4cm}
	\includegraphics[width=0.5\textwidth]{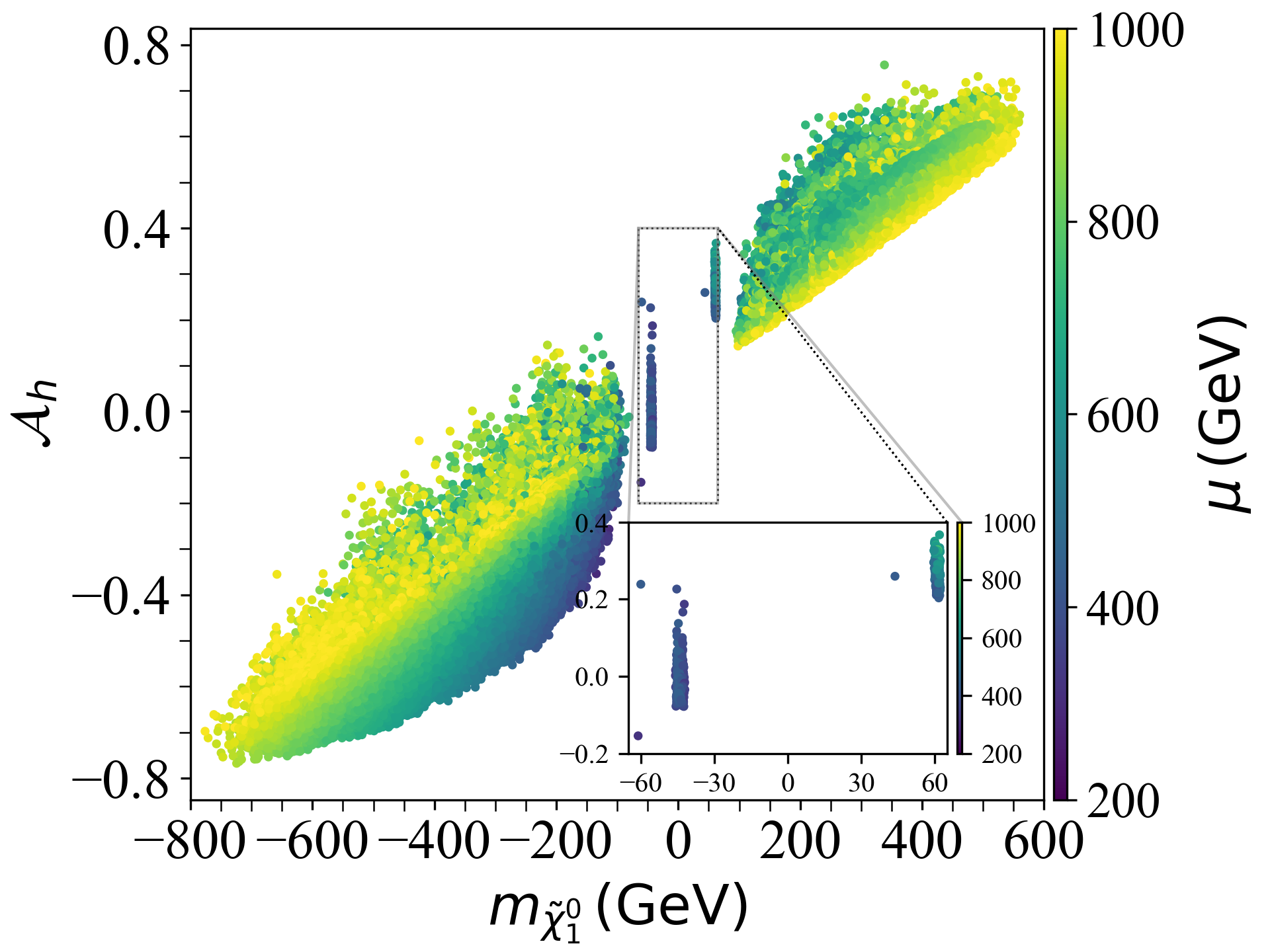}
	\caption{\label{fig2}
Projection of the refined samples onto the
 ${\cal{A}}_h-\sigma^{SI}_{\tilde\chi^0_1 - p}$ plane, where the
  colors indicate the contributions from H to the
SI cross-section ${\cal{A}}_H$, and onto the $m_{\tilde\chi^0_1}-{\cal{A}}_h$ plane, where the colors indicate the value of the higgsino mass $\mu$.}
\end{figure}

\begin{figure}[t]
	\centering
	\includegraphics[width=0.5\textwidth]{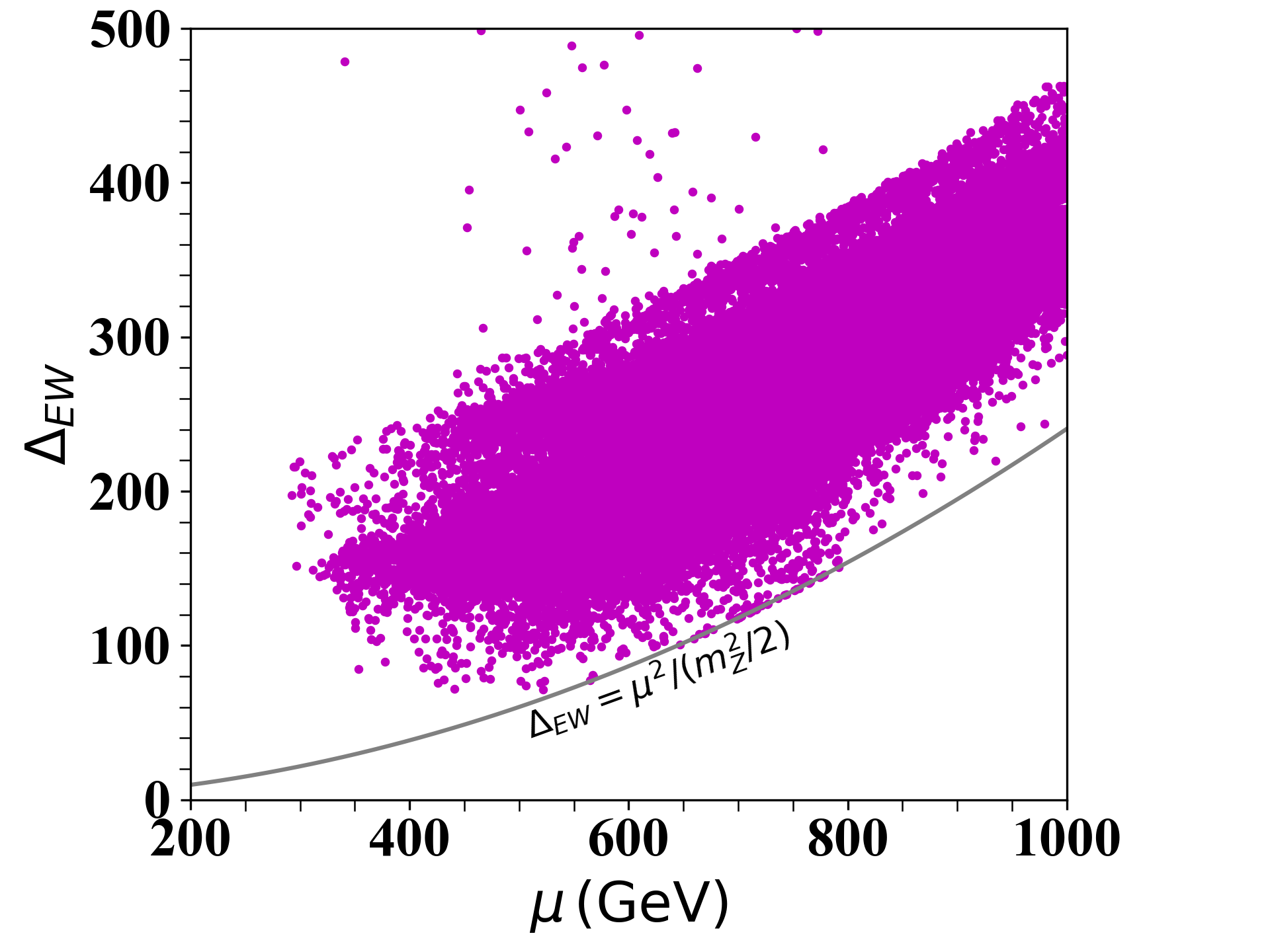}%\hspace{-0.4cm}
	\includegraphics[width=0.5\textwidth]{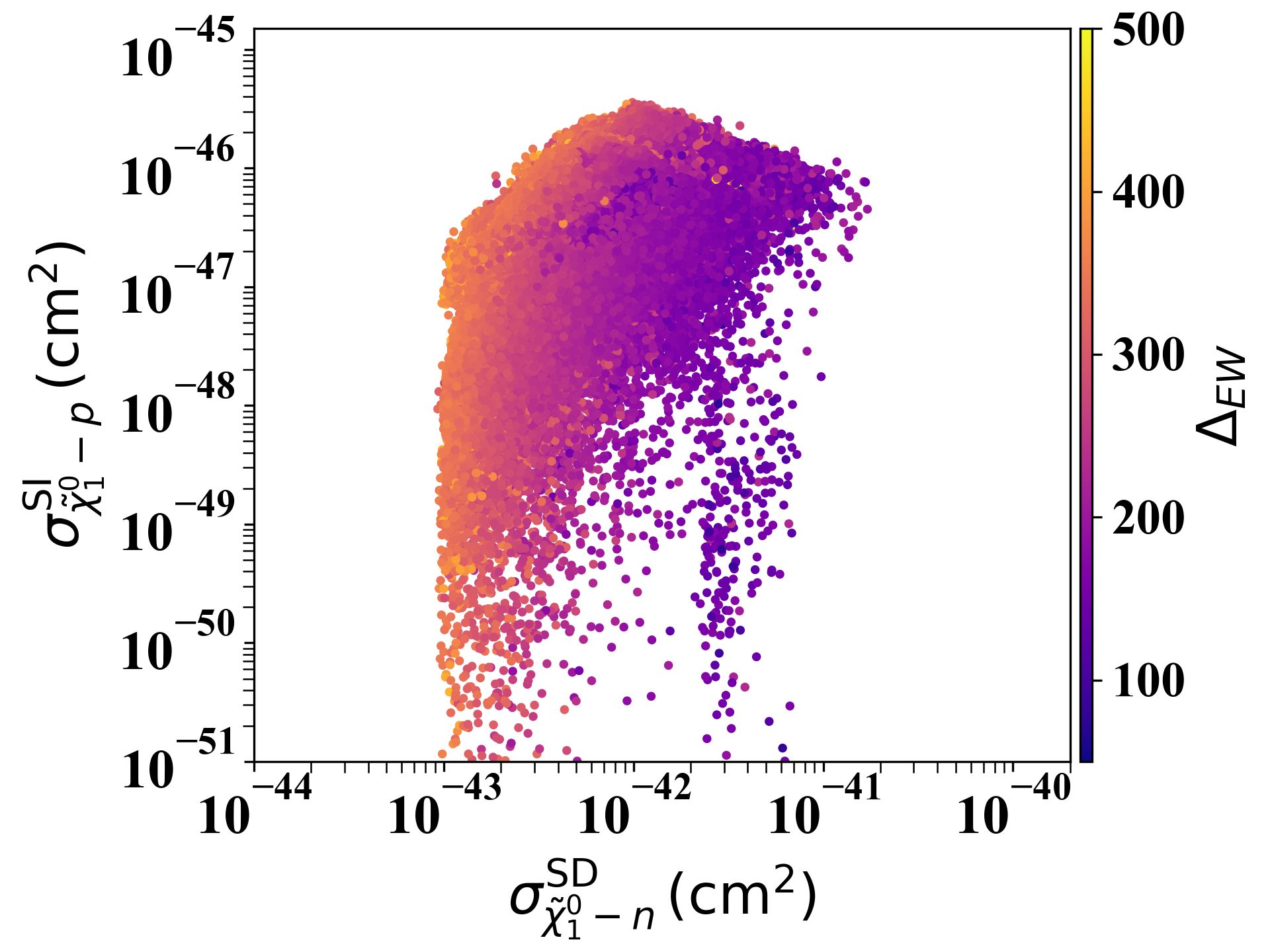}
	\caption{\label{fig3}
Projection of the refined samples onto the $\mu-\Delta_{EW}$ plane and $\sigma^{SD}_{\tilde\chi^0_1 -n}-\sigma^{SI}_{\tilde\chi^0_1 - p}$ plane with the colors indicating the values of $\Delta_{EW}$.}
\end{figure}

The acquired samples were refined using the following criteria: the observed DM relic abundance within $\pm 10\%$ of the measured central value is $\Omega h^2 = 0.12$ ( i.e., $0.108 \leq \Omega h^2 \leq 0.132$)~\cite{Planck:2018vyg}, and $N^2_{11}>0.5$ to guarantee that the LSP is a bino-like neutralino. Then, we projected the refined samples on the corresponding planes of Fig.~\ref{fig1}, Fig.~\ref{fig2} and Fig.~\ref{fig3} .

%During the scan, the likelihood function to guide the
%process considers the experimental constraints, including the consistency of $h$'s properties with the LHC Higgs data at the $95\%$ confidence level (C.L.)~\cite{HS2020uwn},
%collider searches for extra Higgs bosons~\cite{HB2020pkv}, $\pm10\%$ at approximately the central value of the DM relic density from the Planck-2018 data~\cite{Planck:2018vyg} ,
%$90\%$ C.L. upper bounds of the PandaX-4T experiment on the SI DM-nucleon scattering~\cite{PandaX-4T:2021bab} and XENON-1T experiment on the SD scattering~\cite{Aprile:2019dbj},
%$2\sigma$ bounds on the branching ratios of $B \to X_s \gamma$ and $B_s \to \mu^+ \mu^-$~\cite{Tanabashi:2018oca},
%and vacuum stability of the scalar potential, which consists of the Higgs fields and the last two generations of the slepton fields~\cite{Camargo-Molina:2013qva,Camargo-Molina:2014pwa}.

In Fig.~\ref{fig1},
we projected the surviving samples on the $ m_{\tilde\chi^0_1}-\sigma^{SI}_{\tilde\chi^0_1 - p}$ plane and $m_{\tilde\chi^0_1}-\sigma^{SD}_{\tilde\chi^0_1 -n}$ plane , where the colors indicate the value of the higgsino mass $\mu$.
Fig.~\ref{fig1} shows that $\sigma^{SD}_{\tilde\chi^0_1 - n}$ is only related to $\mu$
and will be suppressed by a large $\mu$, which is consistent with the analysis based on Eq.~(\ref{sd1}),
i.e., $\sigma^{SD}_{\tilde\chi^0_1 - n}$ is proportional to $\frac{m_Z^2 v^2 }{\mu^4}$. The current LZ experiment constraint on $\sigma^{SD}_{\tilde\chi^0_1 - n}$ requires that $\mu$ is greater than $370~{\rm GeV}$.
However, the distribution of the $\sigma^{SI}_{\tilde\chi^0_1 - p}$ values is relatively complex. Although large $\mu$ can suppress $\sigma^{SI}_{\tilde\chi^0_1 - p}$,
 $\sigma^{SI}_{\tilde\chi^0_1 - p}$ can also be small for small $\mu$. Based on Eq.~(\ref{si1}), $\sigma^{SI}_{\tilde\chi^0_1 - p}$ should be related to a combination of $M_1$, $\mu$, $\tan\beta$, and $m_H$.

To find the combination that will make $\sigma^{SI}_{\tilde\chi^0_1 - p}$  satisfy the latest experimental limit, as presented in Fig.~\ref{fig2},
we projected the surviving samples onto the
 ${\cal{A}}_h-\sigma^{SI}_{\tilde\chi^0_1 - p}$ plane,
  where the colors indicate the contributions from H to the
SI cross-section ${\cal{A}}_H$, and onto the $m_{\tilde\chi^0_1}-{\cal{A}}_h$ plane, where the colors indicate the value of the higgsino mass $\mu$. The contribution of the heavy Higgs (H) to the SI cross-section for most samples was suppressed below 0.14 by $m_H$.
  Fig.~\ref{fig2} shows that compared with the contributions from H (${\cal{A}}_H$) and $v^2/\mu^2$ to the
SI cross-section,  the contribution of h (${\cal{A}}_h$) played a dominant role.

In Fig.~\ref{fig3}, we projected the samples onto the $\mu-\Delta_{EW}$ plane and $\sigma^{SD}_{\tilde\chi^0_1 -n}-\sigma^{SI}_{\tilde\chi^0_1 - p}$ plane with the colors indicating the values of EWFT ($\Delta_{EW}$). Fig.~\ref{fig3} shows that the samples were fine-tuned  with large $\Delta_{EW} > 70$($\Delta_{EW}^{-1} = 1.4\%$EWFT). According to the above discussion, the reason is that the latest  LZ experiment requires $\mu> 370 {~\rm GeV}$. However, low $\Delta_{EW}$ solutions are only possible for low values of $\mu$ ~\cite{Baer:2012cf,Baer:2013bba}. As shown in the second figure of Fig.~\ref{fig3}, $\Delta_{EW}$ of the samples allowed by the LZ experiment on the SI scattering cross-section may be relatively low due to the existence of the ``generalized blind spot'' . Because the values of $\mu^2/(m_Z^2/2)$ are the lower bounds of $\Delta_{EW}$, we will use $\mu^2/(m_Z^2/2)$ to measure $\Delta_{EW}$ hereafter.

According to DM annihilation mechanisms, we divided the refined samples into four categories to discuss in detail.

\begin{enumerate}
  \item Type-I samples: $m_{\tilde\chi^0_1} \approx -\frac{1}{2}m_Z$ ,
  $5  < \tan \beta <58 $,
  $337 {~\rm GeV} < \mu <462 {~\rm GeV}$,$100 {~\rm GeV} < M_2 <1206 {~\rm GeV}$. ${\tilde\chi^0_1}$ is mainly annihilated to
   $d \bar{d}$ by exchanging a resonant Z-boson in the s-channel to obtain its measured relic density. According to Eq.~(\ref{Z-funnel0}), the corresponding annihilation cross-section in this area is approximately
   \begin{equation}\label{Z-funnel}
    \langle\sigma v\rangle^{d\bar d,Z}_{x_F}\simeq
    (2.5\times10^{-26}\frac{{\rm cm}^3}{s})\left[\frac{2.2\times10^{-3}\times C_{\tilde\chi^0_1 \tilde\chi^0_1 Z}}{\Delta_Z}\right]^2\left(\frac{m_{\tilde\chi^0_1}}{46{\rm ~GeV}}\right)^{-4}.
   \end{equation}

   From Fig.~\ref{fig1},
   $\sigma^{SI}_{\tilde\chi^0_1 - p}$ can also satisfy the direct detection constraints with small $\mu$ values,
    but $\sigma^{SD}_{\tilde\chi^0_1 - n}$  cannot.
     As discussed above,
     $\sigma^{SI}_{\tilde\chi^0_1 - p}$ was proportional to $({\cal{A}}_h+{\cal{A}}_H)$ and suppressed by $\mu^2$, so it even vanishes under the arranged limits of $M_1$, $\mu$, $\tan\beta$, and $m_H$. Fig.~\ref{fig2} shows that  $|{\cal{A}}_h| < 0.1$, which corresponds to ${M_{1}}/{\mu}$ and $\tan\beta $ having opposite signs, e.g., suppresses the SI cross section.
     However, $\sigma^{SD}_{\tilde\chi^0_1 - n}$ is only proportional to $\frac{m_Z^2 v^2 }{\mu^4}$.
     According to Eq. (\ref{sd1}), the LZ direct detection constraint on the SD cross section requires $\mu > 500 {~\rm GeV}$  for these samples,  which corresponds to $\Delta_{EW}^{-1}=1.7\%$.
    Further, according to Eqs. (\ref{bino_xxz}) and (\ref{Z-funnel}), $  \mu =500 {~\rm GeV }$  corresponds
    to $C_{\tilde\chi^0_1 \tilde\chi^0_1 Z} = 3\times10^{-3}$, i.e., to predict the correct relic density, $\Delta_Z =6.6\times10^{-6}$ which implies a tuning of ${\cal{O}}(0.0001\%)$.
    A larger $\mu$ corresponds to a more severe fine tuning, so $\mu$ cannot be very large in this case.

  \item Type-II samples: $m_{\tilde\chi^0_1} \simeq \frac{1}{2}m_h$ ,
  $7 < \tan \beta <33 $, $406 {~\rm GeV} < \mu <776 {~\rm GeV}$, $100 {~\rm GeV} < M_2 <994 {~\rm GeV}$.
${\tilde\chi^0_1}$ is mainly annihilated  to $b \bar{b}$ through the s-channel exchange of a resonant SM-like Higgs boson h to obtain its measured relic density. According to Eq. (\ref{h-funnel0}),  the corresponding annihilation cross-section is approximated to
   \begin{equation}\label{h-funnel}
    \langle\sigma v\rangle^{b\bar b,h}_{x_F}\simeq
    (2.5\times10^{-26}\frac{{\rm cm}^3}{s})\left[\frac{0.2\times C_{\tilde\chi^0_1 \tilde\chi^0_1 h}}{\Delta_h}\right]^2\left(\frac{m_{\tilde\chi^0_1}}{62{~\rm GeV}}\right)^{2}.
  \end{equation}

    Compared with the Type-I samples, the $\mu$ value increased in the case.
    For most samples, $\mu > 500{~\rm GeV}$ so that  $\sigma^{SD}_{\tilde\chi^0_1 - n}$ can satisfy the direct detection constraints, but $\sigma^{SI}_{\tilde\chi^0_1 - p}$ cannot.
     From Fig. \ref{fig2}, the range of ${\cal{A}}_h $ is (0.2,0.37), and
     the elevating effect of $ ({\cal{A}}_h+{\cal{A}}_H)$ on $\sigma^{SI}_{\tilde\chi^0_1 - p}$
     has an advantage over the suppressing effect of $\mu^2$. At this time, $M_{1}/\mu$ and $\tan\beta $ have identical signs.
     For $\tan \beta =10(30)$, according to Eq. (\ref{si1}),
     the direct detection constraints on the SI cross-section increased $\mu$ to above $ {1500 {~\rm GeV}}$ ($ {1200 {~\rm GeV}}$), which
     corresponded to $\Delta_{EW}^{-1}=0.2\%$($\Delta_{EW}^{-1}=0.3\%$).  According to Eq. (\ref{bino_xxh}), for $\mu= {1500 {~\rm GeV} }({1200 {~\rm GeV} })$,
 $C_{\tilde\chi^0_1 \tilde\chi^0_1 h}$$ = 3(4)\times10^{-3}$, i.e., the right relic density requires $\Delta_h =6(8)\times10^{-4}$ which corresponds to a tuning of ${\cal{O}}(0.01\%)$.

  \item Type-III  and Type-IV samples: $m_{\tilde\chi^0_1} \in (-800,-100){~\rm GeV}~ \bigcup~ (100,550){~\rm GeV}$,
  $5  < \tan \beta <60 $,  $300 {~\rm GeV} < \mu <1000 {~\rm GeV}$,$100 {~\rm GeV} < M_2 <1420 {~\rm GeV}$, and ${\tilde\chi^0_1}$ is mainly co-annihilated with wino-dominated electroweakinos ( ${\tilde\chi^0_2}$ or ${\tilde\chi^\pm_1}$ ) (Type-III samples) or sleptons ( ${\tilde\mu_L}$ or ${\tilde\mu_R}$ ) (Type-IV samples) to achieve the measured density.  Here, we used $\Delta_{M_2} = (M_2 - |m_{\tilde{\chi}_1^0}|)/|m_{\tilde{\chi}_1^0}|$, and
  $\Delta_{m_{\tilde{l}}} = (M_{\tilde\mu_{L} or \tilde\mu_{R}}- |m_{\tilde{\chi}_1^0}|)/|m_{\tilde{\chi}_1^0}|$ to parameterize the mass splitting between  ${\tilde{\chi}_1^0}$ and wino-dominated electroweakinos or sleptons. From Fig.~\ref{fig2}, we can conclude that
  \begin{itemize}
    \item the cases with $-400 {~\rm GeV} < m_{\tilde\chi^0_1} <-100{~\rm GeV}$ have the ``generalized blind spot'', so $\sigma^{SI}_{\tilde\chi^0_1 - p}$  can be small for small $\mu$. However, the latest  LZ experiment requires $\mu> 370 {~\rm GeV}$. The right relic density requires $|\Delta_{M_2}|<0.13$ or $|\Delta_{m_{\tilde{l}}}|<0.18$.
    \item for the samples with $100 {~\rm GeV} < m_{\tilde\chi^0_1} <550{~\rm GeV}$, the contributions from h and H on $\sigma^{SI}_{\tilde\chi^0_1 - p}$ reinforced each other so that  $\sigma^{SI}_{\tilde\chi^0_1 - p}$ was relatively large. The latest LZ experiment increased $\mu$ to above $582{~\rm GeV}$ which corresponds to $\Delta_{EW}^{-1}=1\%$.  The right relic density requires $|\Delta_{M_2}|<0.07$ or $|\Delta_{m_{\tilde{l}}}|<0.07$.
    \item for the region with $-800 {~\rm GeV} < m_{\tilde\chi^0_1} <-400{~\rm GeV}$, although $M_{1}/\mu$ and $\tan\beta $ have opposite signs, the values of  $|{\cal{A}}_h|$ were too large so that $\sigma^{SI}_{\tilde\chi^0_1 - p}$ remained large for large $\mu$. The latest  LZ experiment also increased $\mu$ to above $600{~\rm GeV}$, and the right relic density is obtained when $|\Delta_{M_2}|<0.07$ or $|\Delta_{m_{\tilde{l}}}|<0.07$. Briefly, the characteristics of this area are similar to the cases with  $100 {~\rm GeV} < m_{\tilde\chi^0_1} <550{~\rm GeV}$. Moreover, these regions of the parameter space will shrink or disappear with further improvement of sensitivity in future experiments.
  \end{itemize}
\end{enumerate}
%\begin{figure}[t]
%	\centering
%	\includegraphics[width=0.45\textwidth]{mDM_deltam2.png}%\hspace{-0.3cm}
%	\includegraphics[width=0.45\textwidth]{mDM_deltamsl.png}
%	\caption{\label{fig3}
%		Projected the surviving samples onto $m_{\tilde\chi^0_1}- \Delta_{M_2}$ plane  and
% $m_{\tilde\chi^0_1}- \Delta_{m_{\tilde{l}}}$ plane. }
%\end{figure}

\begin{table}[h]
\centering
\caption{\label{Table1}Values of $\mu$ and $\Delta_{EW}$ for four types of samples after and before considering the LZ experiment. $\times$ and $\surd$ indicate that the corresponding experimental limitations cannot and can be satisfied, respectively. }

\vspace{0.2cm}

\resizebox{1.0 \textwidth}{!}{
\begin{tabular}{cc|c|c|c|c|c|c}
\hline
\hline
 \multicolumn{2}{c|}{\multirow{3}{*}{Sample type}}  & \multicolumn{2}{c|}{\multirow{1}{*}{ Before}}
                 & \multicolumn{4}{c}{{After}} \\
\cline{3-8}
 &  & \multirow{2}{*}{$\mu (\rm GeV)$}& \multirow{2}{*}{$\Delta_{EW}$} & \multicolumn{2}{c|}{\multirow{1}{*}{SI}} &\multicolumn{2}{c}{\multirow{1}{*}{SD}} \\
\cline{5-8}
 & &&& $\mu (\rm GeV)$ & $\Delta_{EW}$ & $\mu (\rm GeV)$ & $\Delta_{EW}$ \\
\hline
\multicolumn{2}{c|}{Type-I} & (337,462) & (75,250)& (340,462) & (75,250)& $\times $& $\times$\\
\hline
\multicolumn{2}{c|}{Type-II}  & (406,776) & (72,345) & $\times$ &$\times$ &(445,776) & (78,345)\\
\hline
\multirow{3}{*}{Type-III or IV}& $M_1<-400 {\rm GeV}$& (574,1000) & (117,452) &(607,1000) &(123,452)& $\surd$& $\surd$\\
\cline{2-8}
\multirow{3}{*}{}&$-400{\rm GeV}\lesssim M_1 \lesssim -100 {\rm GeV}$ &(300,1000) & (71,500) & (300,1000) & (80,500)& (370,1000)&(71,500)\\
\cline{2-8}
\multirow{3}{*}{}&$M_1 \gtrsim 100{\rm GeV}$ &(487,1000) & (122,463) & (582,1000) & (174,463)&$\surd$ &$\surd$\\
\hline
\hline
\end{tabular}}
\end{table}

Table~\ref{Table1} summarizes the above discussion of different samples. $\Delta_{EW}$ is the values presented in Fig.\ref{fig3}. When the LZ experiment was considered, the lower bounds of the higgsino mass $\mu$ and $\Delta_{EW}$ were elevated, and $\mu$ should exceed $600~{\rm GeV}$, $370~{\rm GeV}$, $340~{\rm GeV}$, $440~{\rm GeV}$ and $580~{\rm GeV}$  for the cases of $M_1 < - 400~{\rm GeV}$, $- 400~{\rm GeV} \lesssim  M_1 \lesssim - 100~{\rm GeV}$, $M_1 \simeq - m_Z/2$, $M_1 \simeq m_h/2$, and $M_1 \gtrsim 100~{\rm GeV}$, respectively. In particular, $\mu$ was more tightly limited for the case of $M_1 > 0$ than for  $M_1 < 0$, when $|M_1|$ was fixed. The reason is that, under the approximation $m_{\tilde{\chi}_1^0} \simeq M_1 $, a negative $M_1$ can lead to the cancelation between different contributions to the SI DM-nucleon scattering cross-section in Eq.~(\ref{si1}). These improved bounds of $\mu$ imply a tuning of ${\cal{O}}(1\%)$ to predict the $Z$-boson mass and simultaneously worsen the naturalness of the $Z$- or $h$-mediated resonant annihilations to achieve the measured DM relic density. When considering the restrictions from  the LHC experiment, the lower bounds of $\mu$ improved by approximately $100~{\rm GeV}$, and the upper bound of $M_1$ was approximately 570 GeV~\cite{He:2023lgi}, so the surviving region greatly decreased that will be exacerbated when future DM DD experiments fail to detect the signs of DM, which implies larger unnaturalness of the MSSM.

\section{Conclusion}
\label{sec:Conclusion}
In this study, the LZ experiment recently released its first results in the direct search for DM,
where the sensitivities to the SI and SD cross-sections of DM-nucleon scattering
 reached approximately $6.0 \times 10^{-48}~{\rm cm^2}$ and $1.0 \times 10^{-42}~{\rm cm^2}$, respectively, for the DM mass of approximately $30~{\rm GeV}$~\cite{LZ:2022ufs}.
These unprecedented precision values strongly limit the DM coupling to SM particles that are determined by the SUSY parameters. Considering this strong experimental limitation,
the DM phenomenology and unnaturalness related to DM physics in the MSSM were discussed in detail using approximate analytical formulas.
It was found that under the limit of the latest dark matter experiment,
the unnaturalness associated with DM in the MSSM is embodied in the large higgsino mass $\mu$ elevated by the latest LZ experiment requirement. For example,
after considering the LZ experiment, the lower bounds of the higgsino mass $\mu$ and $\Delta_{EW}$ were elevated, and $\mu$ had to be larger than $600~{\rm GeV}$, $370~{\rm GeV}$, $340~{\rm GeV}$, $440~{\rm GeV}$ and $580~{\rm GeV}$  for the cases of $M_1 < - 400~{\rm GeV}$, $- 400~{\rm GeV} \lesssim  M_1 \lesssim - 100~{\rm GeV}$, $M_1 \simeq - m_Z/2$, $M_1 \simeq m_h/2$, and $M_1 \gtrsim 100~{\rm GeV}$, respectively. In particular, $\mu$ was more tightly limited for the case of $M_1 > 0$ than for $M_1 < 0$, when $|M_1|$ was fixed. These improved bounds of $\mu$ implied a tuning of ${\cal{O}}(1\%)$ to predict the $Z$-boson mass and simultaneously worsen the naturalness of the $Z$- and $h$-mediated resonant annihilations to achieve the correct DM relic density. If the LHC experiment is included, the surviving region greatly decreases which could be exacerbated if future DM DD experiments fail to detect the signs of DM, thus implying larger unnaturalness of the MSSM.

\section*{Acknowledgments}

We sincerely thank Prof. Junjie Cao for numerous helpful discussions and his great effort to improve the manuscript. We thank LetPub for its linguistic assistance during the preparation of this manuscript. This work is supported by the Research and Practice Program on Research Teaching Reform in Undergraduate Colleges and Universities in Henan Province under Grant No.2022SYJXLX100, as well as the Research and Practice Program on Higher Education Teaching Reform in Henan Province under
Grant No. 2024SJGLX0529.
%
%%%%%%%%%%%%%%%%%%%%%%%%%%%%%%%%%%%%%%%%%%%%%%%%%%%%%%

%
\end{document}